\documentclass[a4paper,11pt]{article}

\usepackage[nolist]{acronym}
\usepackage[english]{babel}
\usepackage{booktabs}
\usepackage{braket}
\usepackage[tableposition=top]{caption}
\usepackage{comment}
\usepackage{dsfont}
\usepackage{float}
\usepackage[T1]{fontenc}
\usepackage{graphicx}
\usepackage[italic]{hepnames}
\usepackage[utf8]{inputenc}
\usepackage{longtable}
\usepackage{mathrsfs}
\usepackage{mathtools}
\usepackage{pdfpages}
\usepackage{physics}
\usepackage{placeins}
\usepackage{pos}
\usepackage{setspace}
\usepackage{siunitx}
\usepackage{slashed}
\usepackage{subcaption}
\usepackage{tensor}
\usepackage[colorinlistoftodos]{todonotes}
\usepackage{url}

\usepackage{hyperref}
\usepackage{cleveref}
\usepackage{watermark}
\usepackage[firstpageonly=true,angle=270,scale=0.2, fontsize=20pt,color=gray, pos={22cm,7cm}]{draftwatermark}
\SetWatermarkText{MS-TP-22-55}
\SetWatermarkScale{1}
\definecolor{nordic_blue}{HTML}{5E81AC}
\definecolor{nordic_legend}{HTML}{D8DEE9}
\definecolor{Black}{HTML}{000000}
\numberwithin{equation}{section}

\begin{acronym}
	\acro{cls}[CLS]{Coordinated Lattice Simulations}
	\acro{ddhmc}[DD-HMC]{domain decomposition hybrid Monte Carlo}
	\acro{dfg}[DFG]{Deutsche Forschungsgemeinschaft}
	\acro{fve}[FVE]{finite volume effect}
	\acro{gevp}[GEVP]{Generalised Eigenvalue Problem}
	\acro{hpc}[HPC]{high performance computing}
	\acro{ib}[IB]{isospin breaking}
	\acro{ptbc}[P$\theta$BC]{partially twisted boundary condition}
	\acro{palmaii}[PALMA II]{Parallel Linux System for M\"{u}nster Users II}
	\acro{rhs}[RHS]{right-hand side}
	\acro{wwu}[WWU]{Westf\"{a}lische Wilhelms-Universit\"{a}t M\"{u}nster}
\end{acronym}

\newcommand{\corr}{\ensuremath{\mathcal{C}}}
\newcommand{\decay}{\ensuremath{\HepProcess{\Pgya\to\PD\APD}}}

\newcommand{\pfd}{\ensuremath{\corr_{\HepProcess{\PD\to\PD}}}}
\newcommand{\pfdd}{\ensuremath{\corr_{\HepProcess{\PD\APD\to\PD\APD}}}}
\newcommand{\pfdds}{\ensuremath{\corr_{\HepProcess{\PD\APD\to\PD\APD},~\text{box}}}}
\newcommand{\pfddt}{\ensuremath{\corr_{\HepProcess{\PD\APD\to\PD\APD},~\text{direct}}}}
\newcommand{\pfpdd}{\ensuremath{\corr_{\HepProcess{\Pgya\to\PD\APD}}}}
\newcommand{\pfpp}{\ensuremath{\corr_{\HepProcess{\Pgya\to\Pgya}}}}

\title{Direct access to hadronic decay parameters with twisted boundary conditions}

\author[a]{B. Blossier}
\author[b]{J. Heitger}
\author*[b]{J. Neuendorf}
\author[a]{T. San José}

\affiliation[a]{
	Laboratoire de Physique des 2 Infinis Irène Joliot-Curie,
	CNRS/IN2P3,\\
	Université Paris-Saclay,
	F-91405 Orsay Cedex,
	France}
\affiliation[b]{
	Institut für Theoretische Physik,\\
	Westfälische Wilhelms-Universität,
	Wilhelm-Klemm-Straße 9, 48149 Münster, Germany}

\emailAdd{blossier@ijclab.in2p3.fr}
\emailAdd{heitger@wwu.de}
\emailAdd{jan.neuendorf@wwu.de}
\emailAdd{san-jose-perez@ijclab.in2p3.fr}

\abstract{
Our exploratory study looks for direct access to the resonant hadronic transition amplitude without resorting to the L\"{u}scher formalism.
We study the decay \decay~by applying \acp{ptbc} to the quenched charm quark,	circumventing possible problems with final state interactions.
If successful, we could compute the dependence of the transition amplitude on the charm-quark mass and test the predictions made by phenomenological quark-pair-creation models.
Finally, we study if and to what extent an extraction	of the excited state \Pgya is necessary for this analysis.
}

\FullConference{%
 The 39th International Symposium on Lattice Field Theory (LATTICE2022)\\
  8th-13th August 2022\\
  Rheinische Friedrich-Wilhelms-Universität Bonn, Germany\\
}

\begin{document}
\renewcommand{\qty}{\SI}

\onehalfspacing
	\maketitle
	\section{Introduction}
 
Investigating hadronic decays from lattice QCD is notoriously challenging. Indeed, analytical properties of decay amplitudes, like the poles in the complex plane that can be identified as resonances, are not straightforward to recover from correlation functions in Euclidean space. However, L\"{u}scher's formalism and its various generalisations have made a breakthrough in the study of processes like the \Prho meson decay, or scatterings of meson-meson, meson-baryon and baryon-baryon states.
As those approaches relate the decay parameters to the spectrum of interacting particles in finite volume, they  require computations in several physical volumes to identify singular points on phase shift curves. As a consequence, the numerical work can be costly.

McNeile et al.~\cite{McNeile:2000xx} proposed a different method, which gives direct access to the decay matrix elements under the condition that the kinematical configuration is close to the threshold.
In our exploratory work, we combine the approach in \cite{McNeile:2000xx} with the use of \acp{ptbc} so that the threshold condition can be established off the momenta quantisation in finite volume. 

\section{Particle decay under investigation}

\begin{figure}
\centering	
\includegraphics[width=0.55\columnwidth]{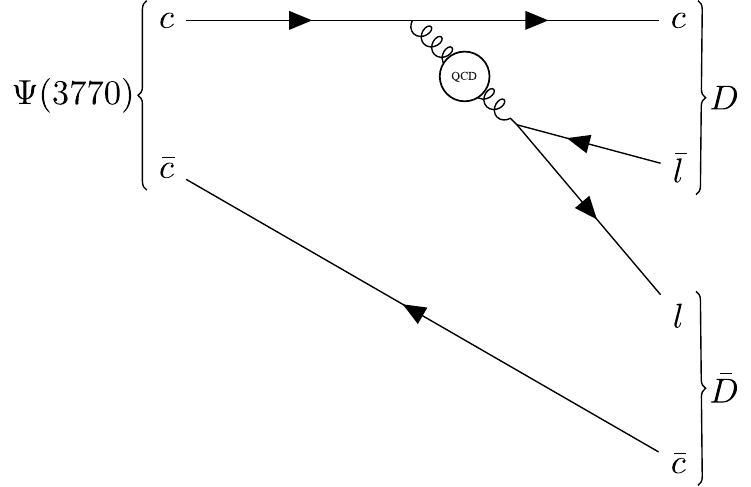}
\caption{$\decay$ decay Feynman diagram.}
\label{fig:decay}
\centering
\includegraphics[width=0.85\columnwidth]{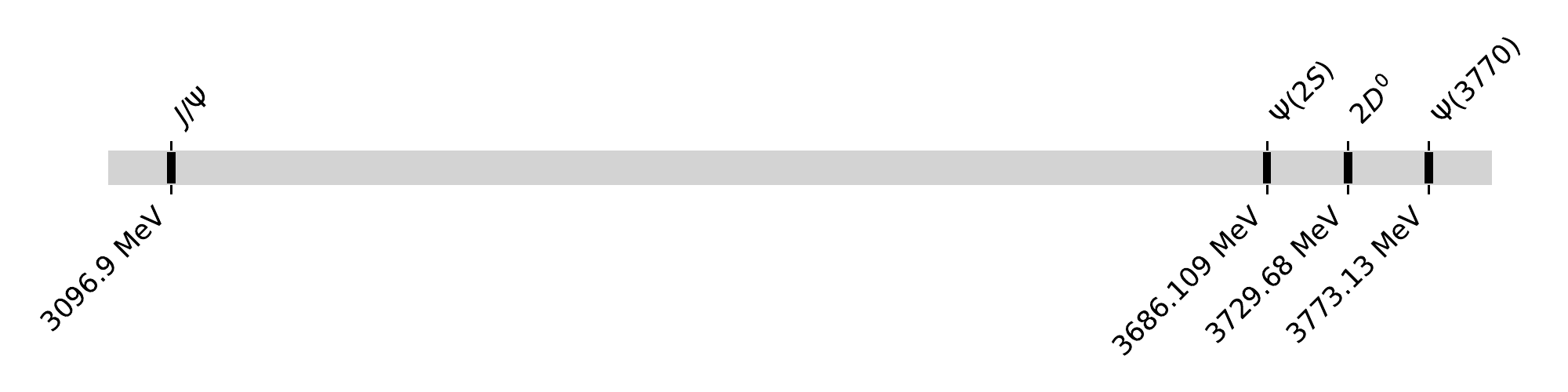}
\caption{Experimental spectrum of vector charmonium and $D\bar{D}$ \cite{ParticleDataGroup:2022pth}.}
\label{fig:spectrum}
\end{figure}

To apply the proposal in \cite{McNeile:2000xx}, we study the charmonium state \Pgya, which is a vector particle with quantum numbers $\text{I}^{\text{G}}(\text{J}^{\text{PC}})=0^{-}(1^{--})$ and mass $M_{\Pgya} = \qty{3773.7(4)}{\mega\eV}$ \cite{ParticleDataGroup:2022pth}.
We consider its main decay channel, \decay, depicted in \cref{fig:decay}, with $(\Gamma_{\PDz}+\Gamma_{\PDpm})/\Gamma_{\text{total}}>90\%$. Since we do not account for \ac{ib}, we cannot distinguish between the neutral and charged \PD mesons in the final state.
Then, the braket that we aim to compute is
\begin{equation}
	x\equiv\braket{\PD\APD\,}{\,\Pgya}.
\end{equation}
This decay is ideal for probing \cite{McNeile:2000xx} because it occurs nearly at the threshold in an experiment, $M_{\Pgya}-2M_{\PDz} \sim \qty{44}{\mega\eV}$ \cite{ParticleDataGroup:2022pth}. The relevant experimental spectrum for our work appears in \cref{fig:spectrum}. Of course, the unphysical quark masses used in the simulations and a finite lattice spacing can affect the particular values of the masses seen in \cref{fig:spectrum}, and they might even make the $\PD\APD$ state heavier than $\Pgya$.
As the initial and final states, $\Pgya$ and $\PD\APD$, are very close in mass, the twist angle required to fix the kinematical configuration at the threshold is expected to be small.
Moreover, twisted boundary conditions boost each meson of the $\PD\APD$ system in the opposite direction, increasing the system energy while it stays at rest.

Since we study a process with final-state interactions between the \PD and \APD mesons, we may only apply \ac{ptbc} to the quenched charm quark \cite{Sachrajda:2004mi}, keeping finite-volume effects under control.
The decay $\decay$ has been thoroughly studied in experiments, and it offers a nice playground to test the reliability of theoretical frameworks like early effective quark models.
In this project, we concentrate on the $^3P_0$ quark pair creation model \cite{LeYaouanc:1973ldf}, which was fruitful for understanding the dynamics of hadronic meson decay in two mesons \cite{Kokoski:1985is}. It postulates an effective charm-quark mass dependence only through the wave functions that describe each of the three hadrons entering the process, and introduces a universal coupling $\gamma$.
In turn, we compute the matrix element $\braket{H_q \overline{H}_q}{\bar{h}h(1^-,\,n=3)}_{E[H_q \overline{H}_q]=m(\bar{h}h(1^-,\,n=3))}$ on the lattice, and compare the dependence on the mass $m_h$ of the heavy quark $h$.%
\footnote{Here, $\bar{h}h(1^-,\,n=3)$ is the second radial excitation of the vector quarkonium while $H_q$ is the ground-state pseudoscalar heavy-light meson. At $m_h=m_c$, $\bar{h}h(1^-,\,n=3) \equiv \psi(3770)$ and $H_q \equiv D$.}

\section{Methodology}

We consider correlation functions of operators $O_{\bar{h}\bar{q} h q}$ and $O_{\bar{h}h}$, which work as interpolators of the $D\bar{D}$ and $\psi(3770)$ states, respectively. 
\begin{figure}
	\centering
	\includegraphics[width=0.8\textwidth]{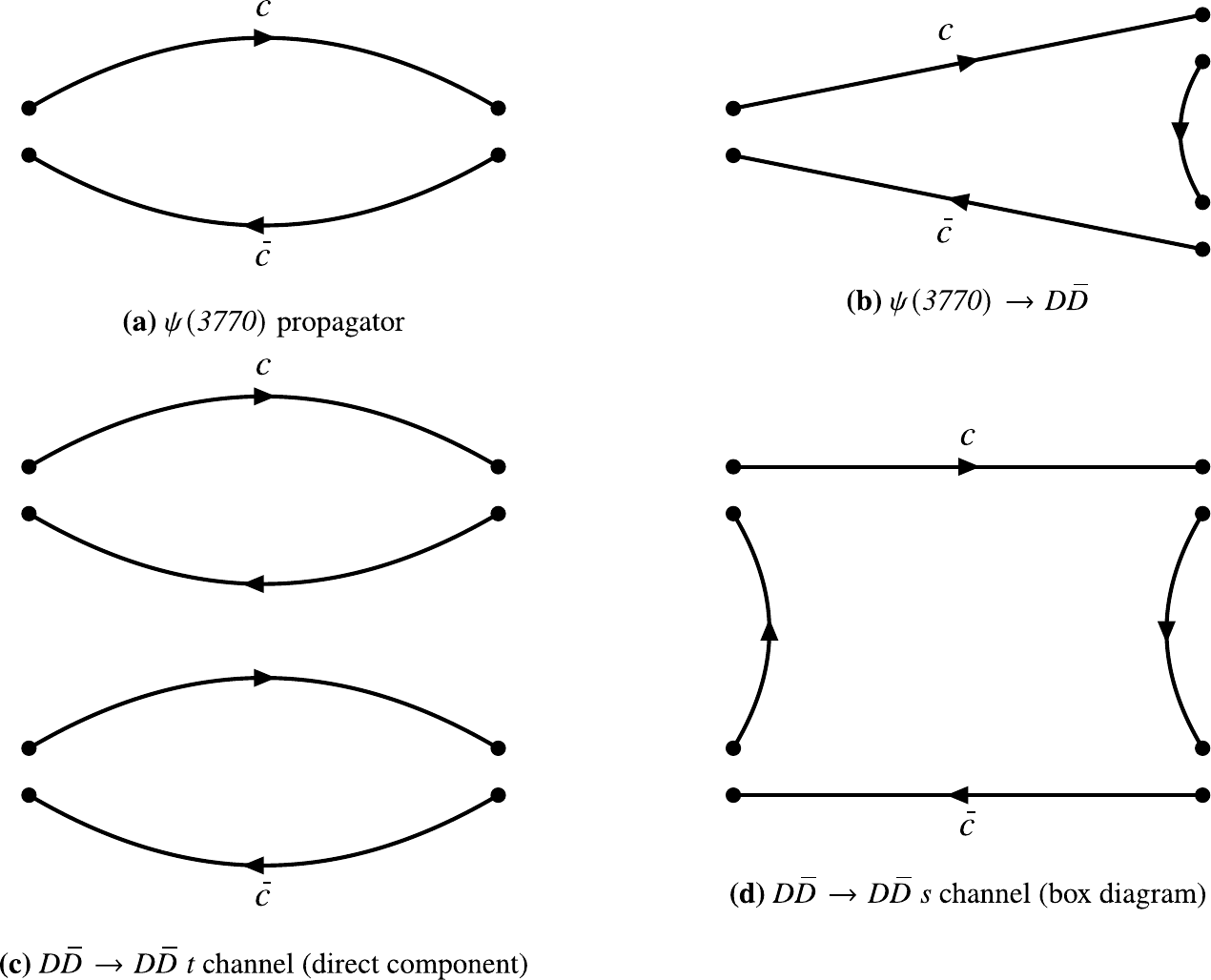}
	\caption{Diagrams needed for our study \cite{McNeile:2002fh}.}
	\label{alldiagrams}
\end{figure}
To extract the matrix element $x$ from the lattice, we need to study the asymptotic behaviour of the correlators whose diagrams are shown in \cref{alldiagrams} \cite{McNeile:2002fh},
\begin{equation}
	\label{eq:asymptotic-behaviour}
	\begin{gathered}
		\pfpp(t)
			\equiv
			\expval{O_{\bar{h}h}(t) O^{\dag}_{\bar{h}h}(0)}
			\longrightarrow
			Z_{\Pgya}^2~e^{-m_{\Pgya}t},
		\\
		\pfdd(t)
			\equiv
			\expval{O_{\bar{h}\bar{q}hq}(t)O^{\dag}_{\bar{h}\bar{q}hq}(0)}
			\longrightarrow
			Z_{\PD\APD}^2~e^{-E_{\PD\APD}t},
		\\
		\pfpdd(t)
			\equiv
			\expval{O_{\bar{h}\bar{q}hq}(t)O^{\dag}_{\bar{h}h}(0)}
			\longrightarrow
			\sum_{t_i} Z_{\Pgya}~e^{-m_{\Pgya}t_i}~x~e^{-E_{\PD\APD}(t-t_i)}~Z_{\PD\APD},
	\end{gathered}
\end{equation}
where the full correlator for $\PD\APD \to \PD\APD$ process is 
\begin{equation}
	\pfdd = \pfd^2 + \pfdds.
\end{equation}

The asymptotic behaviour of $\pfpdd(t)$ can be derived using the transfer matrix formalism.
Calling $\ket{\Pgy(n)}$ the $1^-$ quarkonia states and $\ket{D\bar{D}(n)}$ the $D\bar{D}$ mesons $P$-wave states, both series of states are eigenmodes of the transfer matrix.
Then, if we assume the threshold condition $m_{\Pgy(n=3)} = E_{D\bar{D}(m=1)}$\footnote{$\Pgy(n=3)\equiv \psi(3770), D\overline{D}(m=1) \equiv D\bar{D}$.} and $\mel{\Omega}{O_{\bar{h}h}}{D\bar{D}(n)} = \mel{\Omega}{O_{\bar{h}\bar{q} h q}}{\Pgy(m)}=0$, we have
\begin{equation}
\begin{aligned}
&\expval{O_{\bar{h}\bar{q}h q}(t) O^{\dag}_{\bar{h}h}(0)}
\\
&=
\sum_{0 \leq t_1 < t} \sum_{m, n}
	\mel{\Omega}{O_{\bar{h}\bar{q}h q}}{D\bar{D}(n)}
	~\braket{D\bar{D}(n)}{\psi(m)}
	~\braket{\psi(m)}{\Omega}
	~e^{-m_{\psi(m)} t_1} e^{-E_{D\bar{D}(n)}(t-t_1)}
	+\dots
\\
&=
\sum_{0 \leq t_1 < t}
	\mel{\Omega}{O_{\bar{h}\bar{q} h q}}{D\bar{D}(1)}
	~\braket{D\bar{D}(1)}{\psi(3)}
	~\braket{\psi(3)}{\Omega}
	~e^{-E_{D\bar{D}(1)} t}
	+\dots
\\
&\sim
	\mel{\Omega}{O_{\bar{h}\bar{q}h q}}{D\bar{D}(1)}
	\braket{\psi(3)}{\Omega}
	~x~t~e^{-E_{D\bar{D}(1)}t}.
\end{aligned}
\label{eq2hadto1had}
\end{equation}
To extract $\ket{\Pgya}$ and $m_{\Pgya}$, we solve a \ac{gevp} where we consider several Gaussian smearing levels of the quark field, covariant derivatives and Dirac structures.
Setting $E_{\PD\APD}=m_{\Pgya}$ allows to find the appropriate \ac{ptbc} $h(x+L \vec{e}_i)=e^{i \theta_i L} h(x)$ for the heavy field obeying the threshold condition.\footnote{Note that we have imposed isotropic twisted boundary conditions.}

We may use the asymptotic behaviour in \cref{eq:asymptotic-behaviour} in two different ways to extract $x$ from the lattice \cite{McNeile:2000xx,McNeile:2002fh,McNeile:2002az}. The first method is to consider the decay $\decay$ directly,
\begin{equation}
	\label{eq:tx+const}
	\frac{\pfpdd(t)}{\sqrt{\pfdd(t)~\pfpp(t)}}
		\quad \underset{t/a \gg 1}{\longrightarrow} \quad tx + \text{const},
\end{equation}
where the dominant excited-state contamination appears in the constant term.
The factor
\begin{equation}
	\mel{\Omega}{O_{\bar{h}\bar{q} h q}}{D\bar{D}(1)}
	~\braket{\psi(3)}{\Omega}
	~e^{-E_{D\bar{D}(1)}t}
\end{equation}
of $\pfpdd(t)$ simplifies versus the denominator on the left-hand side of \cref{eq:tx+const}.
The other method is to only use the information from $\PD\APD\to\PD\APD$,
\begin{equation}
	\label{ratio_eq}
	\frac{\pfdds}{\pfddt} \quad \underset{t/a \gg 1}{\longrightarrow}\quad \frac{1}{2} x^2 t^2 + \order{t}.
\end{equation}
Here, it is interesting to study the dependence (if any) of the residual linear term on the twist angle $\theta$ used for the boundary conditions.

\begin{table}[t]
\begin{center}
\begin{tabular}{cccccccccc}
\toprule
id 	& $\beta$ & $(\frac{L}{a})^3\times\frac{T}{a}$ & $a~[\textrm{fm}]$ & $m_{\Ppi}~[\textrm{MeV}]$ & $m_{\pi}L$ & $\#$~cfgs & $\kappa_{\rm \ell}$ & $\kappa_s$ & $\kappa_c$
\\
[0.2cm]
\hline
\\
D5	 & 5.3 & $24^3\times48$ & 0.0653 & 439 & 4.7 & 150 & 0.13625 & 0.135777 & 0.12724 \\
F7 & & $48^3\times96$ & & 268 & 4.3	& t.b.d. & 0.13638 & 0.135730 & 0.12713 \\    
\bottomrule
\hline
\end{tabular} 
\end{center}
\caption{\Ac{cls} $N_f=2$ ensemble parameters. From left to right: Ensemble label, inverse bare coupling $\beta=6/g_0^2$, lattice geometry and spacing, pion mass, $m_{\Ppi}L$, number of configurations in our study, and $\kappa$ values for the bare light, strange and charm quarks (see main text for more details). Moreover, we know that at leading order the \acl{fve} decays exponentially fast with $m_{\Ppi}L$, and $m_{\Ppi}L \geq 4$ reduces significantly its impact. Finally, we plan to extend our analysis to F7 to study the pion mass dependence of our results, but no statistics is available yet.}
\label{tab:cls-ensembles}
\end{table}

\section{Preliminary Findings}

In this section, we show the status of our analysis on ensemble D5, an $N_f=2$ \textit{Coordinated Lattice Simulations} \ac{cls} simulation, whose details appear in \cref{tab:cls-ensembles}. $N_f=2$ \acs{cls} ensembles employ the Wilson plaquette for the gauge action, as well as $\order{a}$-improved Wilson quarks for the fermionic action. The gauge configurations were computed using the \acs{ddhmc} algorithm, the light-quark hopping parameter yields pion masses between $\qty{190}{\mega\eV} < m_{\Ppi} < \qty{630}{\mega\eV}$, the strange-quark mass is fixed to its physical value by setting $m^2_{\PK}/f^2_{\PK}$ and $m^2_{\Ppi}/f^2_{\PK}$ to their physical values, and the charm-quark mass is chosen such that $m_{\PDs} = m_{\PDs,\text{phys}}$. For more details on the ensemble simulations, see \cite{Fritzsch:2012wq,Heitger:2013oaa}.

\begin{figure}
\centering
\includegraphics[width=0.8\columnwidth]{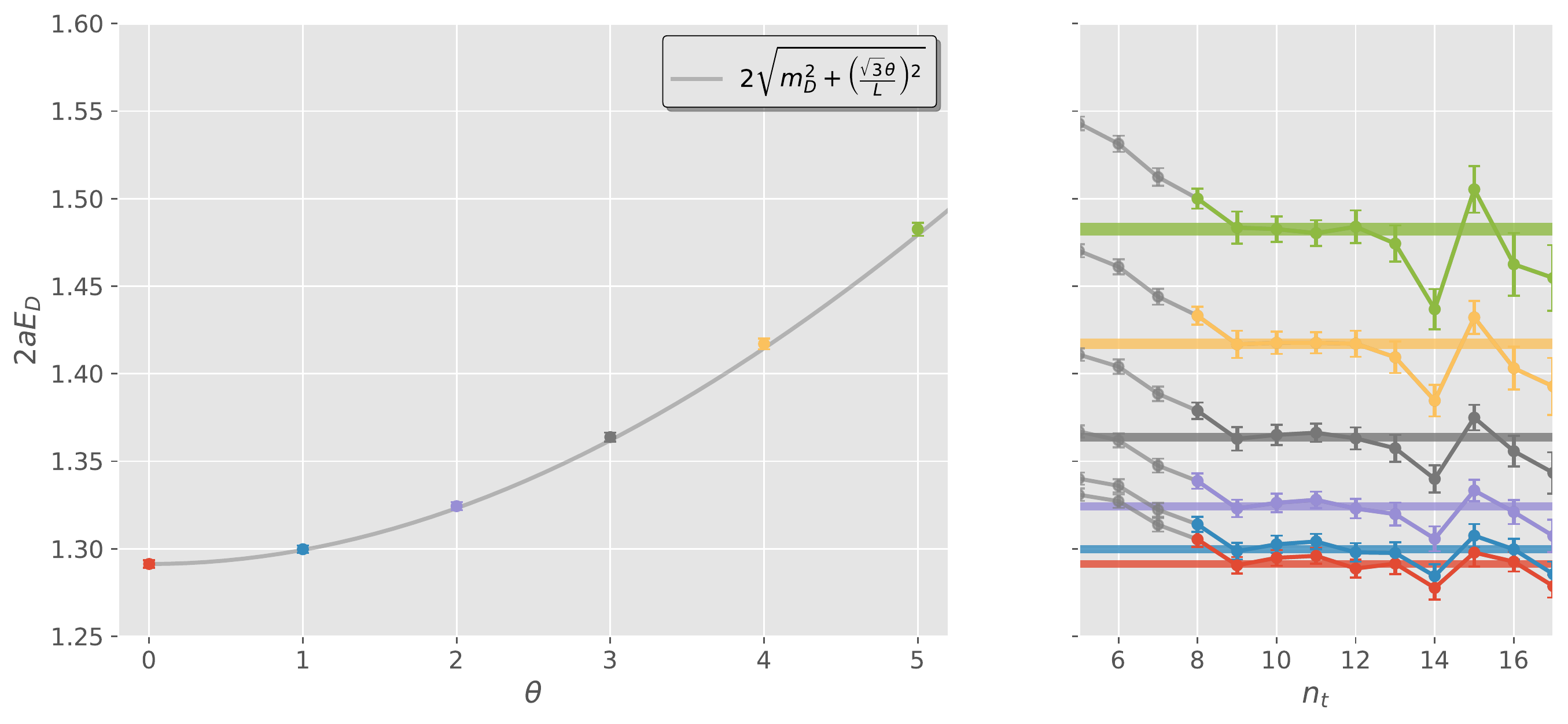}
\caption{Left: Dispersion relation of the $\PD\APD$ system as a function of the twist angle $\theta[1/L]$, assuming $E_{\PD\APD} = 2E_{D}$ and $m_D=E_D(\theta=0)$. Right: $\PD\APD$ effective mass for the various twist angles plotted on the left-hand side as a function of the time-slice.}
\label{fig:dispersion}
\end{figure}

\begin{figure}
\centering
\includegraphics[width=0.8\columnwidth]{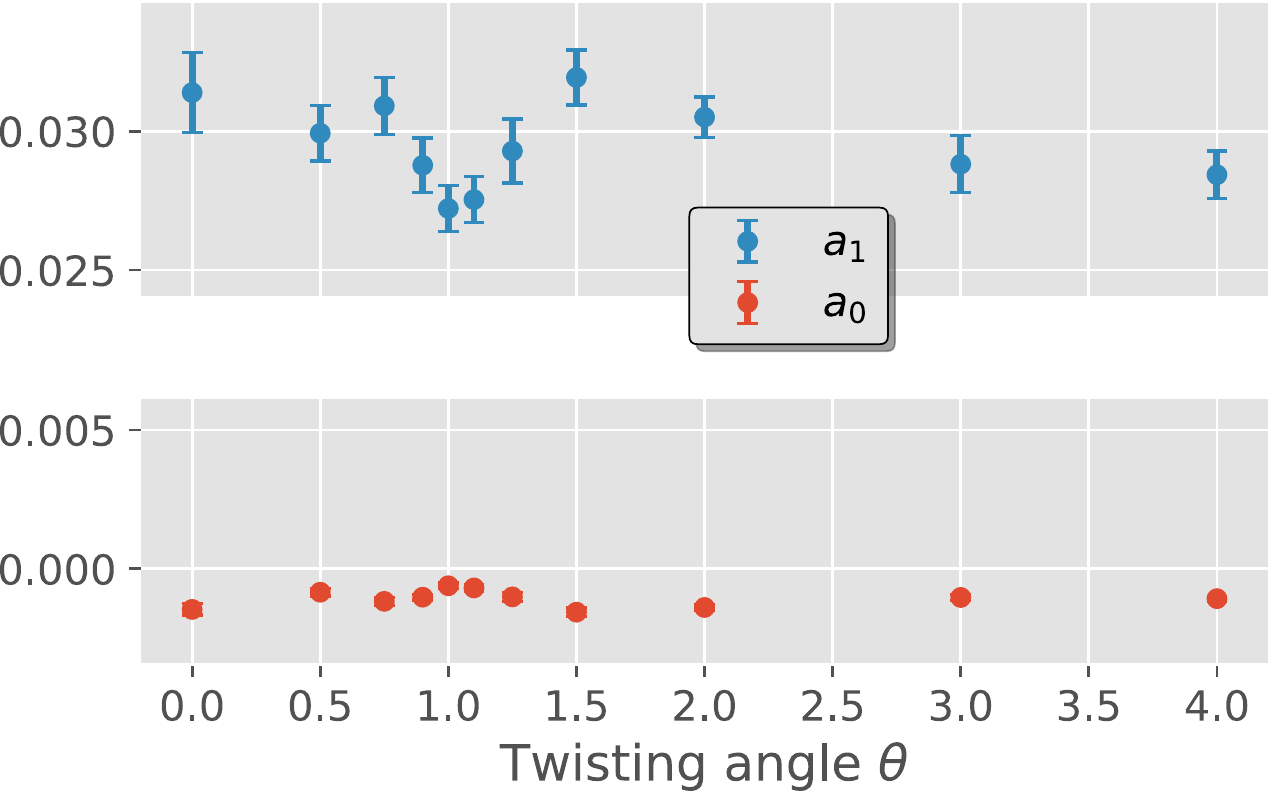}
\caption{Fit parameters $a_0$ and $a_1$ for the model $a_0t + \dfrac{1}{2}a_1^2t^2$ used to fit the ratio $\pfdds/\pfddt$ versus the twist angle $\theta[1/L]$.}
\label{fig:ratio}
\end{figure}

In \cref{fig:dispersion}, we plot the $\PD\APD$ state energy as a function of the \ac{ptbc} used for the charm quark. We observe that the data is well described by the continuum dispersion relation for two free \PD mesons, depicted by the grey line.
Note that at $\theta=0$, corresponding to a $\PD\APD$ system with each meson at rest, the system on ensemble D5 is heavier than its physical value. This is to be expected, due to the unphysical pion mass $m_{\Ppi} = \qty{440}{MeV}$.
So far, our main obstacle has been identifying the $\Pgya$ state from the charmonium spectrum.
We use a basis of operators belonging to the $T_1^{--}$ representation of the H(3) symmetry group, which overlaps with states of quantum numbers $J^{PC}=1^{--}$, $3^{--}$, $4^{--}$,~$\dots$ \cite{Mohler:2012na}, and we employ different levels of Gaussian smearing.
At this stage, we have already computed \cref{ratio_eq} and fitted the function $a_0t + \dfrac{1}{2}a_1^2t^2$ to the result. The matrix element we seek corresponds to $a_1$, which is plotted together with $a_0$ in \cref{fig:ratio} as a function of the twist angle.
As expected, we observe the linear term parameter $a_0$ to be small. Moreover, we observe a certain behaviour of $a_0$ and $a_1$ at $\theta=1$, which we cannot explain at this point. Once we extract $m_{\Pgya}$ and the twist angle at the resonance, we could determine if this behaviour is correlated to the resonance, or not.
For the moment we only include statistical errors estimated using the package \texttt{pyerrors} \cite{Joswig:2022qfe}, which is based on the $\Gamma$-method \cite{Wolff:2003sm}.

\section{Outlook}

Currently, our efforts revolve around an accurate determination of the $\Pgya$ state mass. We are probing different interpolators and smearing levels for the \acs{gevp}.
Afterwards, we will be able to determine the resonant twist angle, fit to \cref{eq:tx+const}, and extract the corresponding decay width. Then, we will extend our calculations to more ensembles, starting with F7 (see \cref{tab:cls-ensembles}), to study the quark mass and lattice spacing dependence of our results. In particular, the quark mass dependence will serve to probe the validity of the $^3P_0$ quark model described in \cite{LeYaouanc:1973ldf}.

\section{Acknowledgements}

Calculations for this project have been performed on the \acs{palmaii} \acs{hpc} cluster at \ac{wwu}. The authors gratefully acknowledge the support of the \ac{wwu}-IT staff. 
This project is supported by Agence Nationale de la Recherche under contract ANR-17-CE31-0019 (B.B. and T.S.) and by the \ac{dfg} through the Research Training Group GRK 2149 \textit{Strong and Weak Interactions -- from Hadrons to Dark Matter} (J.N. and J.H.).
The authors are grateful to their colleagues of the \ac{cls} effort for providing the gauge field ensembles used in this work.

\bibliographystyle{utphys}
\bibliography{references.bib}
	
\end{document}